\documentclass{emulateapj}
\newcommand{\be}{\begin{eqnarray}}
\newcommand{\ee}{\end{eqnarray}}

\newcommand{\nuclei}[2]{\ensuremath{\mathrm{^{#1}#2}}}

\newcommand{\kB}{\ensuremath{k_B}}

\newcommand{\Msolar}{\ensuremath{M_{\odot}}}

\shorttitle{Hydrogen Burning on Magnetar Surfaces} 
\shortauthors{Chang, Arras and Bildsten}

\begin{document}

\title{Hydrogen Burning on Magnetar Surfaces}
\author{Philip Chang\altaffilmark{1},Phil Arras\altaffilmark{2},
and Lars Bildsten\altaffilmark{1,2}}

\altaffiltext{1}
{Department of Physics, Broida Hall, University of California, Santa
Barbara, CA 93106; pchang@physics.ucsb.edu}
\altaffiltext{2}{Kavli Institute for Theoretical Physics, Kohn Hall,
University of California, Santa Barbara, CA 93106;
arras@kitp.ucsb.edu, bildsten@kitp.ucsb.edu}

\begin{abstract}

We compute the rate of diffusive nuclear burning for hydrogen on the
surface of a ``magnetar" (Soft Gamma-Ray Repeater or Anomalous X-Ray
Pulsar).  We find that hydrogen at the photosphere will be burned on
an extremely rapid timescale of hours to years, depending on
composition of the underlying material. Improving on our previous
studies, we explore the effect of a maximally thick ``inert" helium
layer, previously thought to slow down the burning rate. Since
hydrogen diffuses faster in helium than through heavier elements, we
find this helium buffer actually increases the burning rate for
magnetars.  We compute simple analytic scalings of the burning rate
with temperature and magnetic field for a range of core
temperature. We conclude that magnetar photospheres are very unlikely
to contain hydrogen. This motivates theoretical work on heavy element
atmospheres that are needed to measure effective temperature from the
observed thermal emission and constrains models of AXPs that rely on
magnetar cooling through thick light element envelopes.

\end{abstract}


\section{Introduction}

The surface composition of young, isolated neutron stars is not
well constrained by theory.  Since the spectrum is formed in the
outermost $\sim 10^{-17} M_\odot$, it is difficult to predict the
surface composition from supernova simulations, especially in light
of uncertainties in the amount of fallback, nuclear burning in the
nascent star, and subsequent accretion and spallation reactions. On the
observational side, the absence of well-understood atomic spectral lines,
and the extreme physical conditions at the photosphere impede progress
using spectral modeling, though some inferences have been made (
Pavlov, Sanwal \& Zavlin 2002).

Diffusive nuclear burning (DNB) is a mechanism by which the composition
of young neutron star (NS) surface layers can evolve with time (Chang
\& Bildsten 2003; 2004, and Chang, Arras \& Bildsten 2004, hereafter
paper I, II and III respectively). The simple picture of DNB (Chiu \&
Salpeter 1964, Rosen 1968) can be understood as follows (see Fig. 1 of
paper I). Rapid sedimentation ($\sim 1 \,{\rm sec}$) allows H to settle
above heavier elements. While nuclear burning timescales for H at the
surface are far too long to be of interest, a diffusive tail readily
extends to depths at which proton captures occur, increasing the burning
rate by many orders of magnitude.

Anomalous X-Ray Pulsars (AXP's) and Soft Gamma-Ray Repeaters (see
e.g. Woods \& Thompson 2004 for a review), are a class of strongly
magnetized ($B \sim 10^{14-15}{\rm G}$), hot ($T_{\rm bb} \sim
(5-7)\times 10^6{\rm K}$) \footnote{ The relation between $T_e$, the
effective temperature used in our theoretical work, and $T_{\rm bb}$,
the blackbody fit, depends on the composition and magnetic field strength
and is uncertain.} , young ($P/2\dot{P}\sim 10^{3-4}{\rm yr}$), and
slowly rotating ($P\sim 5-10\ {\rm sec}$) neutron stars. We will refer
to them as ``magnetars", following Duncan and Thompson (1992).  In this
Letter, we show that the high temperatures and substantial alteration
of the equation of state (EOS) by superstrong magnetic fields combine
to increase the rate of DNB by many orders of magnitude as compared to
other observed neutron stars. This conclusion is robust for a variety
of heavy proton capturing elements, even when a He layer exists between
H and the proton-capturing element. Consequently, magnetar photospheres
are most likely composed of heavy elements, possibly helium.

\section{ Review of DNB }

The rate at which a column $y_H=\int \rho_H dz$ of H is burned by the 
proton-capturing substrate is (Papers I and II)
\be
\dot{y}_H & \equiv & \frac{y_H}{\tau_{\rm col}}
= \int dz \frac{ n_H m_p}{ \tau_H(n_H,n_{\rm pc},T) },
\label{eq:rate}
\ee 
where $z$ is the depth, $n_H$ and $n_{\rm pc}$ are the number density of
H and proton-capturing substrate, respectively, $T$ is the temperature,
and $\tau_H=(\langle \sigma v \rangle n_{\rm pc})^{-1}$ is the local
lifetime of a proton. The two key ingredients in this formula are the
temperature profile and concentration, $f_H=n_H/n_{\rm pc}$, of H in
the diffusive tail. The temperature profile for strongly magnetized NS
envelopes is reviewed in Ventura \& Potekhin (2002).

The concentration of a trace particle with charge $Z_t$ and mass number
$A_t$ in a background $Z_b$ and $A_b$ depends on
the thermal ($\kB T$), Fermi ($E_F$), and Coulomb interaction
($E_{\rm C}$) energies. In the non-degenerate limit, $\kB T \gg E_F$,
the energy gained by separating the two species across a pressure scale
height is $\sim \kB T$. The resulting concentration profile $f=n_t/n_b$
is a power-law with column (paper I) $f \propto y^\delta$ with exponent
$\delta=A_t(Z_b+1)/A_b-Z_t-1$.  In the degenerate limit, if $Z_t/A_t
\neq Z_b/A_b$, the separation energy is $\sim E_F$ over a scale height,
leading to an exponential concentration profile (paper I)
$f \propto \exp \left[ -(Z_t-Z_bA_t/A_b)E_F/\kB T \right]$.  However,
if $\kB T \ll E_F$ and $Z_t/A_t = Z_b/A_b$, the dominant separation
energy is provided by Coulomb interactions $E_{\rm C} \simeq Z^{5/3}
e^2/a_e$ where the mean electron spacing is $a_e=(4\pi n_e/3)^{-1/3}$.
The resulting concentration profile is (paper III) $f
\propto \exp \left[ -0.9 Z_t(Z_b^{2/3}-Z_t^{2/3}) e^2/a_e \kB T \right]$.
The composition profile drops off much more rapidly in the degenerate
limit (exponential) as compared to the nondegenerate limit (power-law).

For hot NS's ($T_e \ga 10^{6.2}\,{\rm K}$) such as magnetars, DNB is
``diffusion limited" (see paper II), meaning the burning occurs in a
thin layer at which the nuclear burning time becomes comparable to the
downward diffusion time $\tau_H = \tau_{\rm diff}$ breaking the
assumption of diffusive equilibrium (paper I, II).  The hydrogen
concentration profile is rapidly cut off in this region. Cool NS ($T_e
\la 10^{6.2}\,{\rm K}$) have ``nuclear limited" DNB with $\tau_H \la
\tau_{\rm diff}$, even in the burning layer.  In this case, burning
occurs in a degenerate layer where the temperature profile becomes
isothermal and hydrogen fraction is exponentially dropping (paper I).

Paper II studied the effect of a ``standard" pulsar strength
($B=10^{12}{\rm G}$) magnetic field on the compositional structure.
The effect of electron quantization into Landau levels was taken into
account in the opacity and EOS. For $B=10^{12}{\rm G}$, it was found
that the burning rate decreased by more than an order of magnitude.
Most of this decrease is due to
the smaller temperature at fixed column for a given core temperature
$T_c$. In addition, $f_H$ in the burning layer decreased as compared
to the $B=0$ case. The reason, clearly seen in Fig. 4 of Paper II, is
that the electric field above the burning layer is {\it on average}
larger for $B=10^{12}{\rm G}$ than for $B=0$, leading to a smaller
diffusive tail of H. However, this is due to an unlucky coincidence
of the lowest Landau level nearly coinciding with the burning layer,
so that the electric field shows oscillations due to the De Haas-van
Alphen effect. Larger magnetic fields (see the $B=10^{13}{\rm G}$ line)
cause the electric field above the
burning layer to be {\it weaker} than the $B=0$ case, {\it increasing}
$y_H$ in the burning layer. As a result, DNB can be much faster in
magnetars than for standard radio pulsar fields.

\section{Magnetic Field Dependence of DNB}
\label{sec:B}

\begin{figure}[tbp]
%  \epsscale{1.2}
\plotone{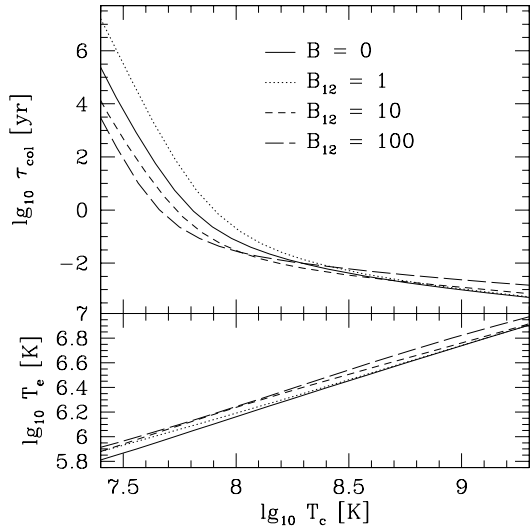}
  \caption{Hydrogen column lifetimes for $p-\nuclei{12}{C}$ as a function of
  $T_c$ for envelopes with vertical magnetic field of different strength
  and $y_H = 10^4\,{\rm g\,cm}^{-2}$.  We show the low field case
  ($B=10^9\,{\rm G}$, thick-solid-line), pulsar field strength
  ($B=10^{12}\,{\rm G}$, dotted-line, and $B=10^{13}\,{\rm G}$,
  short-dashed-line), and magnetar field strength represented by the
  long-dashed line ($B=10^{14}\,{\rm G}$). The bottom panel shows $T_e$
  as a function of $T_c$ for the different field strengths. }
  \label{fig:Te_vs_tau_Bfield3}
\end{figure}

Magnetic fields significantly modify the opacity and EOS when electrons
occupy only the ground Landau level, i.e. $\hbar \omega_{c,e} \ga {\rm
max}(\kB T,E_F)$ where $\hbar\omega_{c,e} = 11.57 B_{12} \,{\rm keV}$
is the cyclotron energy.  There are two main effects.  First, electron
scattering and free-free Rosseland opacities are reduced by a factor $\sim
(\kB T/ \hbar \omega_{c,e})^2$. Hence, for fixed $T_c$, larger $B$ implies
lower temperature at a fixed column (paper II, Potekhin et al. 2003),
and a smaller burning rate.  The photosphere also moves to higher density,
and the $T_e(T_c)$ relation becomes highly dependent on $B$ and $\theta$,
the angle between the field and normal.  Second, the gas becomes more
nondegenerate. The ratio of the $B=0$ Fermi energy $E_{F,3D}$ to the
ground Landau level Fermi energy $E_{F,1D}$ is $E_{F,1D}/E_{F,3D} \sim
(E_{F,1D}/\hbar \omega_{c,e})^{2/3} \ll 1$, pushing the point $\kB T =
E_F$ deeper into the envelope.  This dramatically alters the composition
profile, allowing the slower power-law decrease all the way to the burning
layer, increasing the burning rate. In effect, the parameter space for
``diffusion-limited" DNB is made larger.

The opacity change decreases burning while the degeneracy change increases
burning. Which one wins? The answer is that in the high temperature
limit, the burning rate decreases with $B$, and in the low temperature
limit it increases with $B$ (in the large $B$ limit), as we show in 
this section.

We extend the methodology and results of paper II to higher magnetic
fields and core temperatures. We calculate DNB timescales assuming
a vertical magnetic field, $^{12}$C as the proton-capturing element,
a hydrogen column of $y_H=10^4\ {\rm g\ cm^{-2}}$, and a fixed surface
gravity $g=2.43\times 10^{14}\ {\rm cm\ s^{-2}}$. Varying the magnetic
field direction was discussed in Paper II. The ions are assumed
to be an unmagnetized ideal gas.  We use the results of Chabrier \&
Potekhin (1998, 2000) for the electron equation of state and Potekhin et
al. (1999) for the electron conductivity. We ignore Coulomb interactions
in this section for simplicity, but include it in the next (see Paper
III for details).  Radiative opacities are from Potekhin \& Yakovlev
(2001). The composition profile is computed including diffusion and
nuclear burning (see paper II).  The ion-ion diffusion coefficient in
the nondegenerate regime is from Alcock \& Illarionov (1980) and in
the liquid regime from Brown, Bildsten \& Chang (2002). We refer the
interested reader to paper I and II for additional details.

Fig. \ref{fig:Te_vs_tau_Bfield3} shows the DNB timescale, $\tau_{\rm col}
= y_{\rm H}/\dot{y}_{\rm H}$ as a function of $T_c$ for a range of $B$
at fixed $y_H$.  The low field ($B=10^9\,{\rm G}$) case (solid line)
is included as a baseline. In the low $T_c$ limit, $\tau_{\rm col}$
initially increases with $B$ until $B \simeq 10^{12-13}{\rm G}$, after
which it decreases with $B$.  In the high $T_c$ limit, $\tau_{\rm col}$
increases with $B$. The dependence on core temperature is power law
for high $T_c$ and exponential for low $T_c$.
The rate at which hydrogen is depleted from the
photosphere, $y_{\rm ph} \simeq 500\ {\rm g\ cm^{-2}} T_{e,6}^{5/4}
B_{15}$, can be found using the scaling relation $\tau_{\rm col} \propto
y_H^{-5/12}$.  The column lifetime at the photoshere is far shorter than
the magnetar age ($\sim 10^{3-4}{\rm yr}$).

We estimate the burning rate in the high temperature
limit following paper II. Using free-free opacity with electrons
in the ground Landau level and ideal gas pressure from electrons
and ions gives the temperature profile (Ventura \& Potekhin
2001) $T_7=1.1(T_{e,6}^2 y_6/B_{15})^{4/13}$. The diffusion
coefficient for nondegenerate ions
gives a diffusion time $\tau_{\rm diff} \simeq 2.5
\times 10^4{\rm sec} (B_{15}/T_{e,6}^2)T_7^{9/4}$. The lifetime of
a proton to nonresonant capture on C is $\tau_H \simeq 1.4\times
10^{15}{\rm sec} (T_{e,6}^2/B_{15}) T_7^{-23}$, expanded around
$T_7=1$.  Setting $\tau_{\rm diff}=\tau_H$ gives the temperature,
concentration, column, and capture rate in the burning layer.
Placing these results into eq. (\ref{eq:rate}) gives the
rough analytic formula
\be 
\tau_{\rm col} \simeq 0.1{\rm yr}\ \left(\frac{10^4\ {\rm g\ cm^{-2}}}{y_H}
\right)^{5/12} \left( \frac{B}{10^{15}\ {\rm G}} \right)^{1.2}
\left( \frac{10^6\ {\rm K}}{T_e}
\right)^{2.4}
\ee
in the high $T_c$ limit.

In the low temperature nuclear-limited regime, we estimate the
burning rate following the analytic calculation in Paper I.  In this
case, the burning occurs in a degenerate and nearly isothermal region,
so the hydrogen fraction is $f_H \simeq (y_H/y_{\rm deg})^{17/12}
\exp(1 - E_F/2\kB T_c)= (y_H/y_{\rm deg})^{17/12} \exp(1 - (y/y_{\rm
deg})^{2/3})$, where the column at which electrons in the ground Landau
become degenerate is $y_{\rm deg} \simeq 3.9 \times 10^8{\rm g\ cm^{-2}}
B_{15} T_{c,8}^{3/2}$.  The first factor in $f_H$ accounts for the
decrease in the outer non-degenerate envelope, while the second factor
gives the exponential decrease in the degenerate regime. The temperature
is nearly $T_c$, with a small correction scaling as $y^{-1/3}$. Inserting
this into the p-$^{12}$C rate gives $\exp(-137/T_6^{1/3}) \simeq
\exp(-137/T_{c,6}^{1/3}) \exp\left[-(y_{\rm nuc}/y)^{1/3}\right]$, where
the burning rate changes over a column $y_{\rm nuc} \simeq 7.2\times 10^7
{\rm g\ cm^{-2}} B_{15} T_{e,6}^{12} T_{c,8}^{-7}$. 
Using the method of steepest descents (Paper I)
for the integral in eq.\ref{eq:rate},
the dominant scalings for the column lifetime in the low $T_c$ limit are
\be
\tau_{\rm col} & \propto &  B_{15}^{-7/12} y_{H,4}^{-5/12} 
\exp(29.5/T_{c,8}^{1/3}).
\ee
The scaling of $\tau_{\rm col}$ with $B$ can be explained as two factors
of $B$ from the width of the burning layer and the density in the burning
layer, and the decrease of composition in the nondegenerate envelope
$y_{\rm deg}^{-17/12} \propto B_{15}^{-17/12}$.

\section{Heavier p-capturing Elements and a Helium Buffer}

In section \ref{sec:B} we found that the (diffusion-limited)
burning rate for $p-\nuclei{12}{C}$ is fast enough to deplete H from
the photosphere of a magnetars in a few hours. We now repeat the
calculation for heavier elements Mg, Si and Fe.  The Mg and Si rates
are from NACRE (http://pntpm.ulb.ac.be/Nacre/nacre.htm) and Fe from
Schatz (2004, private communication).  In spite of the large Coulomb
barrier for these elements, we find that H is easily consumed. For Mg,
Si and Fe, the column lifetimes for $y_H = 10^4\,{\rm g\,cm}^{-2}$ at
$B=10^{14}\,{\rm G}$ are, respectively, $\tau_{\rm col}=0.005$, $0.05$
and $2$ years for $T_e = 7\times 10^6\,{\rm K}$, $0.01$, $0.1$ and $5$
years for $5\times 10^6\,{\rm K}$, and $0.03$, $0.7$, and $100$ years
for $3\times 10^6\,{\rm K}$.  These timescales are all much shorter than
magnetar spindown ages $\sim 10^4\,{\rm yrs}$.

\begin{figure}[tbp]
%  \epsscale{1.2}
  \plotone{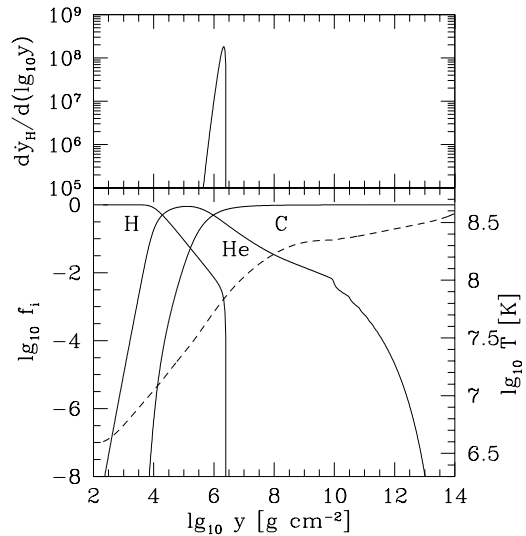}
  \caption{Total H DNB rate including a He buffer
  between the H and C layers. In the lower panel, the solid lines denote
  number fraction of the various species and the dashed line is the
  temperature profile. For $T_e = 4\times 10^6\,{\rm K}$ and
  $y_H = 2 \times 10^4 \,{\rm g\,cm}^{-2}$, the column lifetime
  is $\tau_{\rm col} = 5$ hours.  The total He column is $y_{\rm He}
  \approx 10^{8}\,{\rm g\,cm}^{-2}$ and the B field is
  $B=10^{14}\,{\rm G}$.}
  \label{fig:HHeCMagnetar}
\end{figure}

For sufficiently large Z, the reaction rate is so small that DNB is
nuclear-limited instead of diffusion-limited.  Column lifetimes then
increase exponentially with temperature, as opposed to the slower
power-law increase in the diffusion limited regime.  The transition
between nuclear and diffusion limited burning is given by the
condition $\tau_{\rm diff} = \tau_{\rm H}$ evaluated at the burning
layer. Since the nuclear reaction rate is $\propto \exp(E_0/\kB T)$
where $E_0 \propto Z^{2/3} T^{2/3}$ is the Gamow energy (Clayton 1983),
and $T_e \propto T_c^{1/2}$ (Ventura \& Potekhin 2002), the transition
temperature $T_{e,\rm tr} \propto T_{c,\rm tr}^{1/2} \propto Z$.  For $T_e >
5\times 10^6\,{\rm K}$, DNB occurs only in the diffusion limited, and
hence rapid, regime for elements up to and including Fe.

We now consider the effect of a He buffer between H and the underlying
p-capturing elements.  The H concentration decreases into the He layer
as $f_H \simeq (y/y_H)^{-17/12}$, and the C concentration as $f_C
\simeq (y/y_{\rm He})^2$ in a nondegenerate layer and exponentially in
a degenerate layer, due to Coulomb interactions.  Hence an arbitrarily
thick He buffer would strongly suppress DNB. However, thermal
stability for the 3$\alpha$ reaction sets a maximum He column $y_{\rm
He} \la 10^8\,{\rm g\, cm}^{-2}$ for $T\ga 2 \times 10^8\,{\rm K}$. In
Fig. \ref{fig:HHeCMagnetar}, we plot the H/He/C profile for a magnetar
($B = 10^{14}\,{\rm G}$) with an effective temperature of $T_e =
4\times 10^6\,{\rm K}$.  Because of the high temperatures, even a
maximally thick He buffer makes little difference in the column
lifetime, which is set by the diffusion time down to the burning
layer. A pure H/C envelope at this $T_e$ has $\tau_{\rm col} \simeq
30\ {\rm hours}$. Including the He buffer actually reduces the
lifetime to $\tau_{\rm col} \simeq 5\ {\rm hours}$!  The higher
burning rate is the result of faster diffusion of H through He rather
than C in the diffusion limited regime. To get the rough factor,
consider the diffusion time $\tau_{\rm diff} \propto Z_b^2
A_b^{-1/2}$, where the scaling is found from the diffusion coefficient
in the nondegenerate limit. We find $\tau_{\rm diff}$ is shorter by a
factor of $5$ for He compared to C.  Hence for the high $T_c$ in
magnetars, our estimates for the DNB burning rate are not slowed by
the presence of a He buffer and in the case of diffusion-limited DNB
may be enhanced.

\section{Summary and Conclusions}

We have shown that diffusive nuclear burning removes hydrogen from the
photospheres of magnetars on a timescale much shorter than the age.
This conclusion holds for a range of proton capturing elements up to Fe.
Our new understanding of stratification in Coulomb liquids allows us
to include a maximally thick He buffer, which increases the burning rate
in the diffusion limited regime.

We conclude that magnetar photospheres will not contain hydrogen
unless it can be supplied at a rate
\begin{equation}\label{eq:H-rate}
\dot{M} \ga 2 \times 10^{-16}\,\Msolar\,{\rm yr}^{-1}\left(\frac
{y_H}{10^2\,{\rm g\,cm}^{-2}}\right)\left(\frac {\tau_{\rm col}} {1
{\rm day}}\right)^{-1}.  
\end{equation} 
The burning time is so short that the supply of hydrogen must be
continuous over the timescale of the observations.  The relativistic
outflow expected from magnetars likely precludes accretion of
hydrogen. However, spallation reactions due to magnetospheric currents
may generate fresh hydrogen.

The absence of hydrogen will modify our understanding of both the
thermal evolution and the thermal spectrum of magnetars.  Atomic
physics (e.g. Lai 2001) and spectral modelling of heavy element
atmospheres in $B\sim 10^{14}\,{\rm G}$ fields is relatively
unexplored.  The light atmosphere models of Heyl \& Hernquist (1997),
Ozel (2001) and Ho \& Lai (2001) assume fully ionized H and may be
significantly different if the photospheric material is He or heavy
elements.  Even after $T_e$ has been determined, the question remains
as to whether the data require an additional heat source (e.g. field
decay; Heyl \& Kulkarni 1998) or whether the apparent ``hotness" is
due to a thick layer of low opacity material (i.e.  helium; Heyl \&
Hernquist 1997). Models of AXP's that rely on magnetar cooling with a
H envelope thicker than even a photosphere are difficult to realize
given the rapid timescales we have found for H consumption.  Thermal
stability restricts hot ($T>2 \times 10^8\,{\rm K}$) He envelopes to a
maximum column of $y_{\rm He} \la 10^8\,{\rm g\,cm}^{-2}$.  Since $T_c
\approx 2-3 \times 10^8\,{\rm K}$ (Yakovlev, private communication)
for a non-superfluid NS between the ages of $10^3-10^4$ years, the
effective temperature of such a maximal envelope ($y_{\rm He} =
10^8\,{\rm g\,cm}^{-2}$ with C below) is $T_e \approx 3 \times
10^6\,{\rm K}$ at $B=10^{15}\,{\rm G}$. In comparison a thermally
unstable pure H or pure He envelope would have $T_e \approx 5 \times
10^6\,{\rm K}$ and $T_e \approx 4\times 10^6\,{\rm K}$, respectively.
Therefore, passively cooling neutron stars without internal heat
sources (such as magnetic field decay; Heyl \& Kulkarni 1998 and
references therein; also see Arras, Cumming \& Thompson 2004) fall
short of the observed luminosity of AXPs by a factor of 3 to 30 (Perna
et al. 2001).
  
\acknowledgements 

We thank Hendrik Schatz for providing the cross section for proton
capture onto Fe and Dima Yakovlev for providing NS cooling
models. This work was supported by the National Science Foundation
under PHY 99-07949, and by the Joint Institute for Nuclear
Astrophysics through NSF grant PHY 02-16783.  Support for this work
was provided by NASA through Chandra Award Number GO4-5045C issued by
the Chandra X-ray Observatory Center, which is operated by the SAO for
and on behalf of NASA under contract NAS8-03060.  Phil Arras is an NSF
AAPF fellow.

\end{document}